\documentclass[12pt]{article}

 \usepackage{amsfonts}
 \usepackage{amssymb,amsmath}
 \usepackage{graphicx} \usepackage{epstopdf}
 \newcommand{\crlb}[1]{\label{#1}\\[2pt]}
 
 \newcommand{\crld}[1]{\label{#1}}
 \newcommand{\eela}[1]{\quad\hbox{\scriptsize{#1}}\label{#1}\end{eqnarray}}
 \newcommand{\eelb}[1]{\label{#1}\end{eqnarray}}
 
 \newcommand{\newsecb}[2]{\section{#1}\label{#2}\setcounter{equation}{0}}

 \newcommand{\nolabels} {\def\eel{\eelb}\def\eeql{\eeqlb}  \def\crl{\crlb} \def\newsecl{\newsecb}\def\bibiteml{\bibitem} \def\citel{\cite}\def\labell{\crld}}
\newcommand{\eeqla}[1]{\quad\hbox{\scriptsize{#1}}\label{#1}\end{aligned}\end{equation}}
\newcommand{\eeqlb}[1]{\label{#1}\end{aligned}\end{equation}}

\newcommand\publishversion{\nolabels\setlength{\textheight}{8.3in}\setlength{\oddsidemargin}{0in}
   	 \setlength{\textwidth}{6.3in}\setlength{\topmargin}{-0.2in}}
\def\beq{\begin{equation}\begin{aligned}}		\def\eeq{\end{aligned}\end{equation}}
\def\be{\begin{eqnarray}}  					\def\ee{\end{eqnarray}}		%\be and \ee will become obsolete in due time.
\def\bi#1{\begin{itemize}\item[#1]} 			\def\itm#1{\item[#1]} 			\def\ei{\end{itemize}}
  \def\eqn#1{(\ref{#1})}
   	 \def\fn{\footnote}	  		\def\nm{\nonumber}

		       \def\del{\delta}        \def\lam{\lambda}

    		  	\def\Del{\Delta}   
	    		        		     		\def\vv{\varphi}     
 	 		     	\def\tht{\theta}      	
	     	         
\def\W{\Omega}    		  		\def\dd{{\rm d}}

			\def\ra{\rightarrow}	
\def\bra{\langle} 		\def\ket{\rangle}

\def\fract#1#2{{\textstyle\frac{#1}{#2}}}	 	 
\def\ffract#1#2{\raise .2 em\hbox{$\scriptstyle#1$}\kern-.3em/\kern-.2em\lower .15 em \hbox{$\scriptstyle#2$}}
\def\half{\fract12}					
\def\tl#1{\tilde{#1}}
\def\ex#1{e^{\textstyle#1}} 		\def\qqquad{\qquad\qquad}		
\def\bpmatrix{\begin{pmatrix}} 			\def\epmatrix{\end{pmatrix}}
\def\bmatrix{\begin{matrix}} 			\def\ematrix{\end{matrix}}
\def\bcenter{\begin{center}}			\def\ecenter{\end{center}}
  \def\ds{\displaystyle}
\def\inn{{\mathrm{in}}}  \def\outt{{\mathrm{out}}} 
\def\Pl{{\mathrm{Planck}}}      \def\BH{{\mathrm{BH}}}
\def\widthfig#1#2{\(\hbox{\includegraphics[width=#1]{#2}}\)}
\def\lowerheightfig#1#2#3{\(\raise-#1\hbox{\includegraphics[height=#2]{#3}}\)}
\def\lowerwidthfig#1#2#3{\(\raise-#1\hbox{\includegraphics[width=#2]{#3}}\)}
		 
 \publishversion
\begin{document}

\begin{titlepage}
 \title{\vskip 20mm \LARGE\bf Quantum clones inside black holes}
\author{Gerard 't~Hooft}
\date{\normalsize Institute for Theoretical Physics \\ Utrecht University  \\[10pt]
Princetonplein 5 \\
3584 CC Utrecht \\
 the Netherlands  \\[10pt] internet:
http://www.staff.science.uu.nl/\~{}hooft101/}
 \maketitle

\begin{quotation} \noindent {\large\bf Abstract } \medskip \\
A systematic procedure is proposed for better understanding the evolution
laws of black holes in terms of pure quantum states. We start with the two
opposed regions \(I\) and \(II\) in the Penrose diagram, and study the evolution
of matter in these regions, using the algebra derived earlier from the Shapiro
effect in quantum particles.

Since this spacetime has two distinct asymptotic regions, one must assume
that there is a mechanism that reduces the number of states. In earlier work we
proposed that region \(II\) describes the angular antipodes of region \(I\), the `antipodal
identification', but this eventually leads to contradictions. Our much
simpler proposal is now that all states defined in region \(II\) are exact quantum
clones of those in region \(I\). This indicates more precisely how to restore
unitarity by making all quantum states observable, and in addition suggests
that generalisations towards other black hole structures will be possible.

An apparent complication is that the wave function must evolve with a
purely antisymmetric, imaginary-valued Hamiltonian, but this complication
can be well-understood in a realistic interpretation of quantum mechanics.
\end{quotation}
 Version 4, July, 2022

 \end{titlepage} 

% \tableofcontents

\newsecl{Introduction}{intro}\setcounter{page}{2}
	It has been proposed earlier by this author\,\cite{GtHBH1,GtHfirewall} to regard the Penrose diagram  for the eternal, time independent black hole as a single pure quantum state, on which one then projects the physical excitations  using creation and annihilation operators, so as to obtain the complete spectrum of all pure black hole states that are physically close to the time independent one.
These operators, however, do not commute with the observables describing
the original collapse of the black hole, let alone its final evaporation, and this is
why the effects of collapse and evaporation should not be added to describe the metric
for the background of the quantum effects. We usually proceed by removing collapse and
evaporation, which restricts us to using the metric of an eternal black hole. Sections \ref{pure.sec} and
\ref{KS.sec} describe the basic features needed, such as the Kruskal-Szekeres coordinates. Sections
\ref{Shapiro.sec} and \ref{harm.sec} recapitulate earlier work on the Shapiro effect and its consequences, and on the
algebra that ensues if we apply spherical wave expansions for in- and out going matter.

These findings continue not to be universally accepted or appreciated, but they are
crucial for understanding this paper. At some point it seems as if we are not counting
right. In this paper we make another attempt to be more precise. The main idea of this
paper is treated in section \ref{clones.sec}. How our treatment leads to a unitary description of the
black hole evolution laws is recapitulated in section \ref{unitary.sec}. Some of the problems still left wide
open are mentioned in section \ref{misc.sec} and we phrase our conclusions in section \ref{conc}.
	
\newsecl{The pure, time independent state}{pure.sec}	
Consider the metric of a \emph{static} black hole; We take it to be considerably larger and heavier than the Planck length and mass:
	\be M_\BH\gg 	M_\Pl\ . \ee	
For simplicity this will be the Schwarzschild solution, but generalisation towards the
Reissner-Nordstrom or Kerr-Newman solutions will be straightforward. As is well-known,
this solution does not show a black hole surrounded by a vacuum, but it is accompanied
by a quantum stream of Hawking particles forming a heat bath for the black hole. The
Hawking particles are dominated by a tenuous cloud of particles with masses and energies
far below the Planck value, so that, for the outside world, their direct effects on the total
metric are negligible, though they will be important when considered over long stretches
of time.

In many cases, considering long enough stretches of time, this state is not completely
time-independent. If the Hawking particles are only represented by the ones escaping from
the hole, the black hole mass decreases. A completely static black hole arises only if the
heat bath is carefully tuned to reach perfect time independence. Using standard methods
from quantum mechanics, this can be regarded as a time independent superposition of
three streams of (free) particles: a beam of out going particles (out-particles), a beam of
in going particles (in-particles), and surrounding particles with angular momentum too
high to pass through the angular momentum potential barrier (exterior-particles).

An observer close to the intersection point of the future event horizon and the past
event horizon, de?nes the local energy and momentum density in terms of locally flat
coordinates. (S)he then experiences a local vacuum there, which is a unique vacuum
state, so that, from her point of view, we are dealing with a single pure state. We now
borrow this language to say that, also for the outside observers, this may be regarded as
a single pure state. We claim that any observer will not be able to distinguish a thermal
state from a pure state in thermal equilibrium (a micro-canonical ensemble), so here comes
our first postulate:

	The state where the local observer sees a local minimum of the energy density, i.e. a local
vacuum state, will here be referred to as the Unruh vacuum state\,\cite{Unruh}; it may be regarded
as a pure state. The excited states that we shall use, will in general be time dependent,
small deviations from the Unruh state.
		
\newsecl{The Kruskal-Szekeres coordinates}{KS.sec}
	Now, we consider the analytic continuation of the Schwarzschild metric. Since the local 
observer sees no particles at all, Einstein's equations will require that continuation be carried 
out by assuming strict absence of matter near the black hole. In particular, local observers will 
see no matter crossing future and past horizons. If angular momentum and
electric charge are assumed to be negligible, only the Schwarzschild metric and its analytic
extensions will apply, and therefore it is this spacetime that is the only appropriate one
for describing the stationary black hole in an Unruh heat bath. We have:
\be \dd s^2=\frac 1{1-\fract{2GM}r}\dd r^2-\big(1-\fract {2GM}r\big)\dd t^2+r^2\dd\W^2\ ; \qquad\Big\{ \  \begin{matrix} 
\W&\equiv&(\tht,\vv)\ ,\\ \dd\W&\equiv&(\dd\tht,\,\sin\tht\,\dd\vv)\ .\end{matrix} \eel{Schwmetric.eq}
The ideal continued metric is the Kruskal\,\cite{Kruskal}-Szekeres\,\cite{Szekeres} metric. Replacing \((r,\,t)\) by \((x,\, y)\), with
\beq x\,y&\ = \  \big(\frac r{2GM}-1\big)\ex{r/2GM}\ , \\[5pt]
y/x&\ = \   \ex{t/2GM}\ ,\qquad \eeql{KS.eq}
one derives that the metric is
\be \dd s^2&=& \frac{32(GM)^3}r\,\ex{-r/2GM}\dd x\,\dd y+r^2\dd\W^2\ .  \eel{KSmetric}
This metric is singularity free at the origin, \(x\approx y\approx 0\,,\) which is at \(r\approx 2GM\).	

Now, a natural procedure is to start from here to define time dependent quantum states, by applying field operators,  chosen to be functions of \(x,\,y,\) and the angles \(\W\). Then, however, one obtains too many states, while
quantum information appears to be transported across the horizons. Space-time described by the coordinates \((x,y,\W)\) features \emph{two asymptotic domains rather than one,} simply because every point \((r,t)\) with \(r>2GM\) is associated to one point \(x,y\) with \(x>0\) and \(y>0\) and one point where both \(x\) and \(y\) are negative.

Our proposal in Refs.\, \cite{GtHBH1,GtHfirewall} 
was to impose a constraint on the physical states:\\
All particle states in the domain \(x<0,\ y<0)\) are to be identified with the particles on \((x>0,\,y>0)\) by associating hem to the other's antipodes:  \((\tht,\,\vv)\ra (\pi-\tht,\,\vv+\pi)\).

In this paper we report about our further attempts to implement this idea in practical calculations. We found ourselves forced to \emph{withdraw} the latter additional condition: the antipodal states do not match correctly to the parent states, due to the need to impose \(CPT\) invariance. No antipodal mapping is to be included in the definition of the physical states. There is a much more natural way to limit the physical states: \emph{Only those local
operators that are invariant under the exchange \((x,y)\ \leftrightarrow\ (-x,\,-y)\), are admitted to
define new physical states.} We return to this later (section \ref{clones.sec}).

The danger of omitting the antipodal mapping is that this removes the protection of
our logic against cusp singularities at the origin. We shall have to carefully register what
happens there.

To focus into the business part of the black hole, we go to a coordinate frame that describes space-time very close to the region \(r=2GM\). Write 
	\be x\ra \frac{\sqrt{e/2}} {2GM} u^-\ ;\quad y\ra\frac{\sqrt{e/2}} {2GM}u^+\ ; 	
	\quad
	2GM\equiv R\ ,\quad \fract{1}{4GM}t\equiv\tau .  \eel{nearhor.eq}
In these coordinates, we have	
    	\be\dd s^2\ \ra\  2\dd u^+\dd u^- +R^2\dd\W^2\ ,\eel{nearhormetric.eq}

\newsecl{The Shapiro effect}{Shapiro.sec}
	Of all forces in nature that could explain how black holes process information going in, into information going out again, the Shapiro effect is the most dramatic one. The Shapiro effect is the phenomenon that a radio signal grazing past the Sun is being slowed down by the Sun's gravitational field.
	It is a basic property of the Schwarzschild metric outside the Sun. Any fast moving elementary particles also carries a gravitational field that exerts a Shapiro effect on other particles in its neighbourhood. The effect is computed by subjecting the Schwarzschild metric of a particle at rest, to a strong Lorentz boost. Even if the rest mass of the original particles was negligible (as we shall usually assume), its boosted momentum can easily reach values such that in-particles cause a shift in the positions of the out-particles. 
	
	First consider a small, locally flat region of space-time. In that case, one finds that,, whenever an in-particle meets an out-particle on its way, the out-particle undergoes a shift. If the in-particle has momentum \(p^-\) in the inwards direction, parallel to the coordinate \(u^-\), then the out particle is dragged along in the direction \(u^-\), by an amount \(\del u^-\),  calculated to be:
	\be\del u^-=-4G\,p^- \log|\tl x-\tl x\,'|\ , \eel{dragdelu.eq}
where \(|\tl x-\tl x\,'|\) is the scattering parameter (the transverse separation of the two particles). 

Eq.\eqn{dragdelu.eq}	was first derived in flat space-time\,\cite{Aichelb}, but it is easy to generalise it to the space-time \eqn{nearhor.eq} very close to a black hole horizon\,\cite{GtHDray}. In that case, we replace the transverse
coordinates \(\tl x\)  by the angles \(\W=(\tht,\vv)\):
	\be\del u^-(\W)=8\pi Gf(\W,\,\W') p^-\ ,\eel{horshift.eq}
where \(f(\W,\,\W')\) is a function of the angular separation between the points \(\W\) and \(\W'\), which indicate where the out- and in-particles both puncture through the horizon. The function \(f\)  obeys the angular Laplace equation,
	\be(1-\Del_\W)f(\W,\,\W')=\del^2(\W,\,\W')\ . \eel{angLaplace.eq}.

Thus,  if  \(p^-\)
represents the momentum of an in-particle \emph{added to the system}, at angular position \(\W'\), then
Eq.~\eqn{horshift.eq} tells us that, all out-particles at position \(\W\) are shifted by an amount \(\del u^-(\W)\). 
Our next step is to replace this by 
 the following statement: if \emph{all} in going matter is represented by a function \(p^-(\W')=
 \sum_ip_i^-\del^2(\W_i-\W')\), where \(i\) counts the in-particles, then all out-matter is described by 
 the position function  \(u^-(\W)\), obeying
 \be u^-(\W)=8\pi G\int\dd^2\W' f(\W,\,\W')p^-(\W')\ . \eel{uprelation.eq}
This relation tells us how out-particles are arranged by the in-particles. With Eq.~\eqn{angLaplace.eq}, we have
\be (1-\Del_\W)u^-(\W)=8\pi G p^-(\W)\ . \eel{upalgebra.eq}
It should be understood that the in-particles are defined by their momenta as they cross the future event horizon. 
This (light cone) momentum \(p^-\) increases proportionally to \(\ex{\tau}\), where \(\tau\) if the time coordinate for the outside observer, normalised as in Eq.~\eqn{nearhor.eq}. Similarly, out-particles have momenta \(p^+\)
decreasing as \(\ex{-\tau}\).

The light-cone positions \(u^+\) and \(u^-\) of the in- and the out-particles scale with time, \(\tau\),  exactly as \(p^+\) and \(p^-\) do.

\newsecl{Spherical harmonics}{harm.sec}
It is now instrumental to apply the spherical wave expansion:
\be u^\pm(\W)\equiv\sum_{\ell,m}u^\pm_{\ell m}Y_{\ell m}(\W)\ ,\qquad
p^\pm(\W)\equiv\sum_{\ell,m}p^\pm_{\ell m}\,Y_{\ell m}(\W)\ , \eel{harmexp.eq}
where \(Y_{\ell m}(\W)\), with \(|m|\le \ell\), are the usual spherical harmonic functions
obeying\\ \(\Del_\W Y_{\ell m}(\W)=-\ell(\ell+1)Y_{\ell m}(\W)\ . \) With these, we find that 
Eq.~\eqn{upalgebra.eq} turns into
\be u^-_{\ell m,\, \outt}=\frac{8\pi G}{\ell^2+\ell+1}\,p^-_{\ell m,\, \inn}\ ,\eel{uprel.eq}
where the subscripts in, out, remind the reader that \(u^-\) refers to out-particles and \(p^-\)
refers to in-particles. \(u\) are positions, \(p\) are momenta.

So far, everything was classical.  Now comes quantum mechanics. For the particles, quantum mechanics implies that 
\be [u_i^+,\,p_j^-]=i\del_{ij}\quad (\inn)\ ,\qquad [u^-_i,\,p^+_j]=i\del_{ij}\quad (\outt), \eel{upcomm.eq}
\be\hbox{and consequently,}\qqquad [u^\pm(\W),\,p^\mp(\W')]=i\del^2(\W,\,\W')\ ; \qqquad	\eel{functiecomm.eq}
note that \(u(\W)\) and \(p(\W)\) have been defined as follows:
\be p^\pm(\W)\equiv\sum_i p^\pm_i\,\del^2(\W,\W_i)\ , \qquad u^\pm_i\equiv u^\pm(\W_i)\ , \eel{upOmega.eq}
so that Eqs.~\eqn{functiecomm.eq} follow from \eqn{upcomm.eq}. Defining \(\lam=\ds\frac{8\pi G}{\ell^2+\ell+1}\,, \)\ 
Eq~\eqn{uprel.eq} implies 
\be u^-_{\ell m,\,\outt}=\lam}\,p^-_{\ell m,\,\inn}\ ,\quad \hbox{but also }\quad u^+_{\ell m,\,\inn}=-\lam p^+_{\ell m,\,\outt \ , \eel{uprelplus.eq}
and furthermore, also the variables
\be x=\fract 1{\sqrt 2}(u^++u^-)\quad\hbox{and}\quad p=\fract 1{\lam\sqrt 2}(u^--u^+)\ ,\ee
obey \([\,x,p\,]=i\,,\) while their \(\tau\) dependence is generated by the Hamiltonian 
\be H_{[\tau]}=\half\lam p^2-\fract 1{2\lam}\,x^2\ . \ee
which is that of a repulsive harmonic oscillator, as was noted by Betzios, Gaddam and Papadoulaki\,\cite{BGP.ref}

\newsecl{Clones}{clones.sec} 
\begin{figure} \qquad \widthfig{400pt}{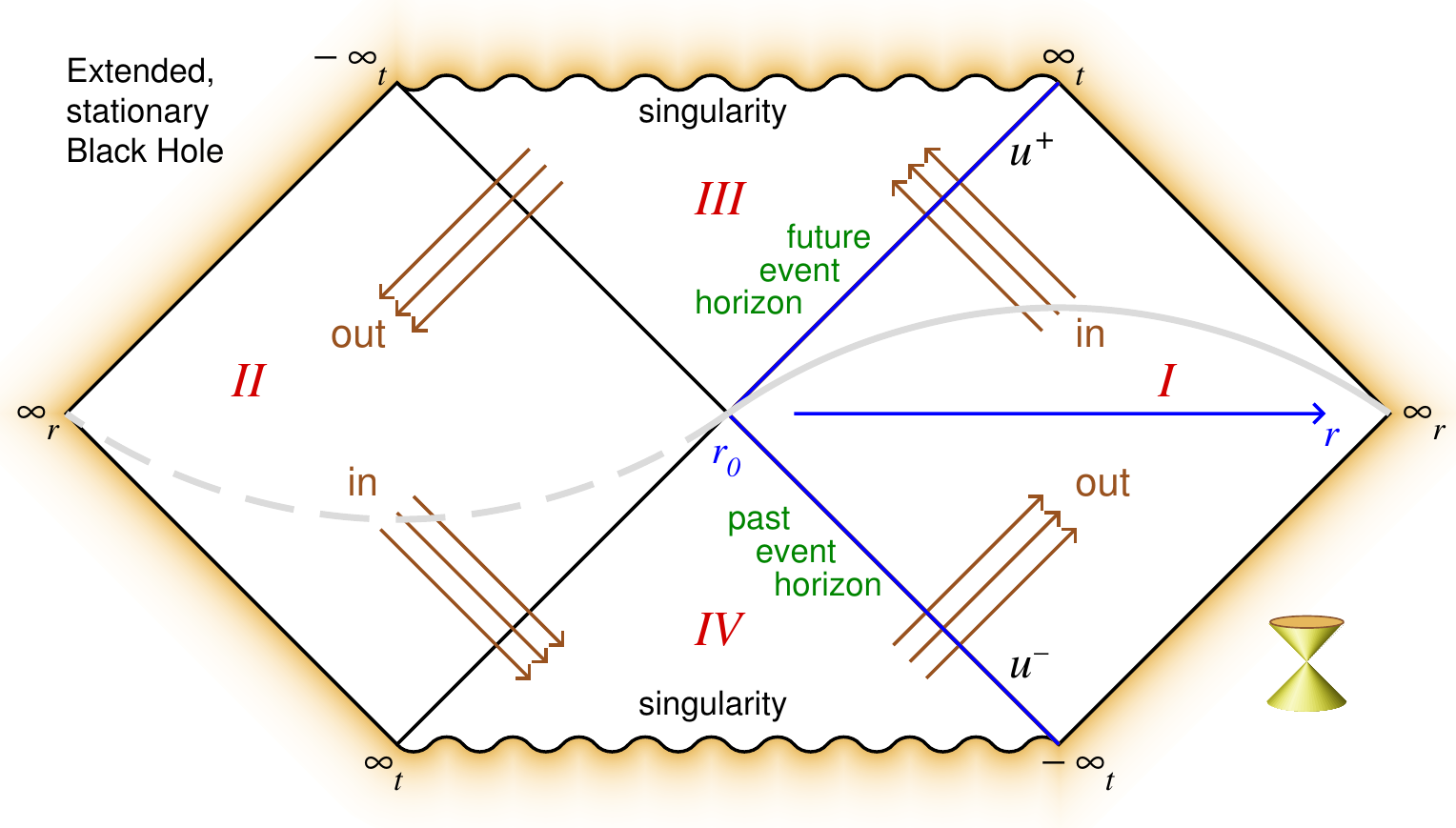} \begin{caption}
{Penrose diagram showing regions \(I \ - \ IV\). In- and out-particles
can punch through a horizon at positive values of \(u^\pm\) (region \(I\)), or negative \(u^\pm\) (region \(II\)).
Grey curve: Cauchy surface. Dashed curve: cloned particles.}
\labell{extpenrose.fig} 
\end{caption}
\end{figure}
	We see that the in-particles punch through the future event horizon either at positive \(u^+\) (region \(I\)) or at negative \(u^+\) (region \(II\)), see Figure \ref{extpenrose.fig}. This ?figure is a Penrose diagram,
obtained by squeezing the coordinates \(u^\pm\) to fit in finite segments, in order to render the entire surrounding universe. The coordinates \(u^\pm\) both span the entire line \([-\infty,\,\infty]\).
	
	Eqs.~\eqn{upcomm.eq}-\eqn{uprelplus.eq} imply that, at every value for
  \((\ell,m)\), the wave function \(\psi_\inn(u^+) \) for the in-particles,  is Fourier transformed to give \(\psi_\outt (u^-)\) for the out-particles. Such a Fourier transform is unitary, so this should yield a perfectly unitary relation between the in- and the out-particles.
  
  However, the particles that escape to region \(II\) are not visible for the observers in region \(I\). This seems to violate unitarity after all. In previous presentations we had assumed that the situation was cured by proposing the particles in region \(II\) to be nothing but the antipodes of particles in region \(I\). The picture seemed to work very well. For instance, the only spots in the Penrose diagram where particles approach their antipodes closely, are the singularities
  in regions \(III\) and \(IV\), which are outside the domains of physical interest and therefore totally harmless.
  
 Yet, we have to withdraw this proposal, mainly for the following reason: written in the Schwarzschild coordinates  \((r,t,\tht,\vv)\), the antipodal mapping used is:\\  \({ }\) \qqquad \((r,t,\tht,\vv)\ \longleftrightarrow \ (r,t,\,\pi-\tht,\,\vv+\pi)\), \\  which implies to a parity reversal \(P\), possibly with a particle-antiparticle  conjugation, \(C\), but no time reversal, \(T\). If, alternatively, we write the transformation in the Kruskal-Szekeres coordinates \((u^+,u^-,\tht,\vv)\), the transformation is \\  \({ }\) \qqquad \((u^+,u^-,\tht,\vv)\ \longleftrightarrow \  (-u^+,-u^-,\pi-\tht,\vv+\pi)\). 
 \\In this notation, we  have  time reversal \(T\) without parity change \(P\).
 
 Both pictures do not follow a \(CPT\) transformation, the only non-trivial reflection operator for which the Standard Model interactions are completely invariant. This would imply that the two regions \(I\) and \(II\)  do no quite evolve the same way, which would be difficult to defend if  the entire Penrose diagram would be assumed to describe just different parts of one and the same space-time.

Rejecting the antipodal mapping brings back to us the danger of a cusp singularity.
This we shall deal with in a different way, see section~\ref{unitary.sec}.

One could suppose that, in the Penrose diagram, simply nothing but a pristine vacuum enters region \(IV\) from region  \(II\)  at time \(\tau=-\infty\). 
In our philosophy, this cannot be right, since region \(II\) describes the same geometry as region \(I\). It cannot be in a vacuum at all. We should use all particles that cross the future event horizon, and Fourier transform them to become the out going particles.
We want the states in region \(II\) to be exactly the ones needed to make the evolution time reversal symmetric and unitary. 

Now let us assume that the particles in region \(II\) are exact quantum clones of the particles in region \(I\), without the antipodal interchange.  At first sight, this seemed to be wrong (which was why it took us so long to get the correct picture!), but let us check.

Quantum clones are defined to be states with exactly the same wave function as some other states. In ordinary circumstances this can practically never arise naturally, but black holes are not natural in this respect. The assumption is now that,  all particles in region \(II\), are described by the same wave functions
as the particles in region \(I\):
	\be\psi(u^+,u^-,\W)=\psi^*(-u^+,-u^-,\W)\ . \eel{clone.eq}
The need for complex conjugation here is not obvious, and it was omitted in the earlier
versions of this paper. But let us temporarily accept Eq.~\eqn{clone.eq}, since \(CPT\) 	invariance does require complex conjugation in this equation.  This is because the time evolution in
the time variable ?\(\tau\) , defined by Eqs. \eqn{KS.eq} and \eqn{nearhor.eq} in regions \(I\) and \(II\) , goes in opposite
directions (see the Cauchy surface, grey curve, in Figure \ref{extpenrose.fig}).
	
	The asterisk in Eq. \eqn{clone.eq} will not be without consequences however, because the
Fourier transformation relating \(u^+\) to \(u^-\)  does not preserve this constraint. What is
going on?
	
Let \(\psi(u^+,0)=\psi_1+i\psi_2\)\,, with \(\psi_1\) and \(\psi_2\) real (particles moving with the speed of light through the future event horizon, have wave functions not depending on \(u^-\)). According to Eq.~\eqn{clone.eq}, \(\psi_1\) is even and \(\psi_2\) is odd in \(u^+\). Then the Fourier transform \(\psi(p)\) defined by
	\be \vv(p)&=&\frac 1{\sqrt 2\pi}
	\int_{-\infty}^\infty\dd x\,\psi(x)\,
	\ex{-ipx}\ , \nm\\[3pt]
	\hbox{obeys}\qquad\vv(p)&=&\sqrt{\frac 2 \pi}
	\int_0^\infty\big(\psi_1(x)\cos px+\psi_2(x)\sin px\big)\ . \eel{phip.eq}
This is a real function but it is \emph{not} even in \(p\), which would be dictated by Eq.~\eqn{clone.eq} if  \(\vv_{1,2}\)
would be real.	

	Therefore, we demand our wave functions to be even in both \(u^+\) and \(u^-\) \emph{and} to be real. The first
condition would be the statement that the wave function for negative \(u^+\) should be a
quantum clone of the function for positive \(u^+\). The second condition is less familiar. Can
we impose wave functions to be real? What would their Schr\"odinger equation look like?
	
From a fundamental point of view, demanding a wave function to be real is easy:
just ensure that the Hamiltonian is imaginary, and hence antisymmetric. In fact, we
can repair complex wave functions rather easily: \emph{complex wave functions are pairs of real
wave functions.} Real wave functions that occur in pairs actually describe a system with
an auxiliary binary variable. The binary variable that we usually have helps us define
energy eigen states. In quantum gravity we may still have a conventional Schr\"odinger
equation, but the energy eigen states all come in pairs of positive and negative energy.
This seems to be a healthy starting point for more advanced quantum theories. We leave
this aspect for future investigations. Right now we can make one interesting remark. The
fact that we have no complex wave functions may entail that we have no conserved global
charges in our theory. This was to be expected in a theory of black holes.\fn{All this may imply that, 
close to the Planckian distance scale, the Standard Model interactions have
to be adapted to the use of specially chosen basis states. This is to be left for further investigations.}
	
\newsecl{Unitary evolution}{unitary.sec}
Now we are in a position to formulate the complete evolution law for a Schwarzschild
black hole:
\bi{1)} Start with the Schwarzschild metric, Eq. \eqn{Schwmetric.eq}.
\itm{2)} Write the wave functions  \(\psi_\inn(u^+)\)of the in-particles in terms of the coordinates
\((r,t,\tht,\vv)\). These wave functions must be real-valued.
\itm{3)} Only the part where \(u^+>0\) is physical. Now define the part for  \(u^+<0\) as being
the quantum clone of the wave for positive \(u^+\,: \ \psi_\inn(u^+)=\psi_\inn(-u^+)\,.\)
\itm{4)} Find the wave function \(\psi_\outt(u^-)\) as the Fourier transform of \(\psi_\inn(u^+) \) over the entire
stretch of \(u^+\) values. Since \(\psi_2=0\)\,,   this Fourier transform obeys the same constraints as \(\psi_\inn\).
\itm{5)} As the Hamiltonian must be imaginary and antisymmetric, the wave function in the
entire region \(II\) is a quantum clone of that in region \(I\) .
\itm{6)} Return to the Schwarzschild coordinates.
\ei

This last step is very important. Our wave function is de?ned in such a way that replacing
\((x,y)\) by \((-x,\,-y)\) yields the same function \(\psi\)\,. Therefore, \(\psi\) is also well-defined on the
original Schwarzschild coordinates \((r,t,\tht,\vv)\)\,.

And now we can return to the cusp singularity. Is there a cusp singularity at the
origin? The answer is no. In terms of the Kruskal-Szekeres coordinates our equations
are totally regular as there is no singularity of the metric. Unitarity holds, so the wave
functions are always normalisable. There may be soft singularities as in other solutions
of the wave equations, but these are totally acceptable from physical point of view.

How do we understand unitarity? What happens if the entire system is postulated to
obey the algebra of Section~\ref{Shapiro.sec}, if we solve the equations in terms of the spherical harmonics
of Section \ref{harm.sec}?

We now describe the main result of this paper: \emph{we can write the evolution equations
exclusively in terms of the data in region \(I\) only. The calculation does employ the Fourier
transformation, but in this formalism it fully respects unitarity.} The wave function \(\psi_\inn(u^+) \)
of particles entering at positive values of \(u^+\) while  \(u^-=0\), generates wave functions
\(\psi_\outt(u^-)\) through Eqs.~\eqn{uprelplus.eq}, \eqn{phip.eq}, which now take the form	
	\be \psi_\outt(u^-)=\sqrt{\frac 2{\pi\lam}} \int_0^\infty\psi_\inn(u^+)\dd u^+\,\cos(u^+ u^-/\lam)\ . \eel{unit1}
This is entirely unitary. Black hole information is preserved. 	
	
	To understand how this works in detail, consider the separate evolution equations at
given \((\ell,m)\). Since the algebraic equations for different \((\ell,m)\) commute, we just need
to look at one generic value of \((\ell,m)\) (this is no longer the case at very high values of
\(\ell\), where the expansion should be terminated; thus we keep only those \(\ell\) values whose
contributions commute).
	
	At these values of \(\ell\)\,, the operators \(u^\pm\) and \(p^\pm\) act as one-particle position 
and momentum operators.\fn{The notion of  `particle'  has to be handled with care; we are considering 
operators u and p obeying the algebra of one-particle position and momentum operators, but the spherical 
wave expansion implies
that they are not particles in the physical sense. For the math, this makes no difference.} The clone 
condition means that the wave function \(\psi\) for the in-particle needs to be defined only for the values 
\(u^+>0\), after which we define the wave function to obey:
	\be\psi_\inn(-u^+)=\psi_\inn(u^+)\ . \eel{unit2}
This is a subspace of all wave functions that may be considered, but the constraints are
necessary for self consistence.

We can now ask how this sub space evolves, when applying the algebra of Section 4. As
stated, the Fourier transform is not unitary on half spaces, but let us check this anyway!

The prototypes of the wave functions for the in-particles obeying Eq.  \eqn{unit2}) describe a
particle at  \(u^+=a\,,\ a>0\)\,:
	\be \bra u^+|\psi\ket=\psi_\inn(u^+)=\del(u^+-a)+\del(u^++a) \eel{unit3}
(to be normalised later). Note that \(u^+\) may be positive or negative, but \(a\) is restricted to be positive.

The out-particle will be the Fourier transform of this. We have
	\be {}_\inn\bra p^-|\psi\ket&=&\fract 1{\sqrt {2\pi}}(\ex{iap^-}+\ex{-iap^-})=\sqrt {\fract 2{\pi}}\cos(ap^-)\ , \label{unit4}\\
	p^-=u^-/\lam &\ra&\bra u^-|\psi\ket\ =\ \bra p^-|\psi\ket/\sqrt\lam = \sqrt{\fract 2{\pi\lam}}\cos(a u^-/\lam)\,.\eel{unit5}
This is again\fn{At first sight, one may now allow the wave functions to be complex. Then, however, the Shapiro
shifts \eqn{uprelation.eq} would include shifts in the wrong direction, so that the classical limit would not be exactly as
in General Relativity.} an even function of \(u^-\) ! Therefore both in \(u^+\) space and in \(u^-\) space, we
can omit the negative halfs. The mapping on the half spaces one onto the other is unitary.
We have, normalising things correctly on the half-spaces \(u^\pm >0\)\,,
	\be\bra u_1^\pm|u_2^\pm\ket&\equiv&\del(u_1^\pm-u_2^\pm)\,,\label{unit6}\\
	{}\inn\bra u_1^+|u_2^-\ket_\outt&=&\sqrt{\fract 2{\pi\lam}}\cos(u_1^+u_2^-/\lam)\ , \eel{unit7}
where, again, the subscripts `in' and `out' are merely to remind the reader that the latter
inproduct connects the out-states to the in-states. The reader is invited to check that
Eqs. \eqn{unit6} and \eqn{unit7} are mutually compatible in the half-spaces  \(u^\pm>0\)\,.

Conclusion: by limiting ourselves to the half-spaces \(u^\pm>0\)\,, we get a completely
unitary transition between the in-particles and the out-particles. This way of seeing how
this can happen, by postulating cloning, is new.

In stead of the even functions, we can also choose to use only odd functions. In that
case:
\be \bra u^+_1|u_2^-\ket = \sqrt{\fract 2{\pi\lam}}\sin(u^+_1 u^-_2/\lam)\,. \eel{unit8}
One can fix the signs here to be positive, without losing generality, since all \((\ell,m)\) values
have to be multiplied.

The attentive reader might now ask: wait, if the metric using \(-x\) and \(-y\) as its coordinates,
is assumed to carry exactly the cloned quantum states of the \(+x,\,+y\) coordinates,
should we not simply continue by putting the quantum states directly onto the original
Schwarzschild coordinates \((r,t,\tht,\vv)\)\,?  Why open up to the Kruskal-Szekeres coordinates
at all?

That would be a very good question. We now believe however that there are two
issues when studying black holes: 1, the quantum states --  to which the question applies,
and 2, the evolution law, for which we need the Shapiro delay, and the algebra that
follows from it. So the answer to the question is:  first open up the metric (i.e. go to the
Kruskal-Szekeres coordinates) to display the evolution laws and their algebra as sharply
as possible, then close it (i.e. go back to the Schwarzschild coordinates), to establish the
boundary conditions. At the boundary, we may assume all quantum states to be in their
cloned form, and, as we established in this section, they will stay cloned forever. Knowing
this, we can continue using the Schwarzschild coordinates.

The rules for using quantum mechanics, in terms of the states in region \(II\) only, are
unchanged.

\newsecl{Miscellaneous open problems.}{misc.sec}
\subsection{Towards The Black Hole Equations}
The problem that we wished to solve is how to derive equations that fully determine
the pure quantum states for the out-going particles if we know how the Standard Model
(SM) projected the in-particles along the future event horizon. The SM only determines
how the particles in region \(I\) are projected against the horizon along the line \(u^+>0\).
Now from this work we know how this fixes the cloned in-particles along the other half
of the future event horizon: the line \(u^+<0\)\,. These wave functions are exact mirror
images. This is already an advance, because before, we could have made a wild guess
at the preferred expression for the momentum density there, whereas now we know the
exact quantum states all over the line \(-\infty<u^+<+\infty\)\,.

This quantum state will have a \(p^-\) distribution, or more precisely, the \(T^{--}\) component
of the momentum density on the  \(u^+\) axis, which determines the \(u^-\) distribution of the
out-particles. Unfortunately, the one question that remains open as yet is how to filter
all SM particles out of this expression, assuming that this can be done at all. It looks as
if knowing  \(T^{++}(u^-)\) alone would not be enough.

On the other hand, modi?cations in the SM may be needed in order to describe the
wave functions as real numbers only; conserved global charges are forbidden.

The attentive reader may have put a question mark at the first sentence of this subsection:
\emph{If the particles entering the future event horizon would have any effect on the
particles leaving the past horizon, would this not lead to closed time-like curves? } It's a
mapping from future to past . . . .

This is true, but this only involves physical effects close to the Planck scale. This
is where we expect the `final physical laws'. We may have to prepare for more novel
phenomena in this domain, including modifications in the laws of physics that will repair
the defects in the equations we are using now. The closed time-like curves expected here
will be a bit like harmonic oscillators, which also return periodically to their previous
states, while this is an important ingredient in the equations that discretises the laws.
More discretisation is indeed expected from the holographic principle in Planck scale
physics.

\subsection{Relation to string theory}
String theory may suggest that exact knowledge of the points where the out-particles
punch through the past event horizon will be all that is needed to characterise a physical
state. The horizon at \(r=2G\,M_\BH\)\,, being a two-dimensional surface, resembles the string
world sheet in a mathematical sense. Section \ref{Shapiro.sec} and \ref{harm.sec}  show that the in-particles enter by
way of Dirac delta peaks on the two-sphere of the horizon in exactly the same fashion
as this is considered in string theories, where one uses \emph{vertex insertions.}\cite{sustr} According to
string theories, the SM particles should re-emerge if we take the zero-slope limit. Vertices
in the bulk represent closed strings entering there, and the lightest components correspond
typically to gravitons, as they topologically represent closed string loops.

One might hope that experience in string theory can help us re-create all particles
present in the string zero slope limit, by linking string language with the black hole
language employed here.

\subsection{What happens at \(r=2GM\)?}
In earlier publications we had been concerned about identifying regions \(I\) and \(II\) of
the Penrose diagram. This would just be like a boundary condition at the edges of
a box, by itself a reasonable assumption. But if such boundary conditions would act
upon states that approach each other in coordinate space, one might expect deformations
that could even lead to singularities. This did not seem right, and a beautiful remedy
seemed to be available, the antipodal mapping: identify any point in region \(II\) with
the antipode of that point in region \(I\). The transverse coordinates form a sphere, with
a radius \(r\ge 2GM_\BH\) everywhere in regions \(I\) and \(II\), so the separation between the
identified antipodal points is \(\ge \pi R=2\pi GM_\BH\)\,.  This is more than sufficient to remove
any cusp singularity; only in regions \(III\)  and \(IV\), there are singularities at \(r\downarrow 0\)\,, which
play no role in our considerations.

But the way we phrased the `boundary condition' now, forbids the direct involvement
of the antipodes. We must identify region \(II\) with region \(I\) directly, at the same spot in
the angular coordinates \(\W=(\tht,\vv)\)\,. The antipodes do not exactly match the points of
the original. Think of an astronaut flying in at one point \(\W\) on the sphere. There is no
astronaut at the antipode, so the metric does not exactly match. We cannot postulate
that the antipode is a clone of the point where the astronaut is situated, and consequently
the formalism would break down.

It is important to add that identifying the wave functions and fields on points \((x,y,\W)\)
with those on \((-x,-y,\,\W)\)  must be regarded as a constraint on the wave functions used,
not on the physical equations. The physical equations were identical anyway because they
refer to the same spot \((r,t,\W)\) in the Schwarzschild coordinates, and this suffices to ignore
any danger arising when points approach their clones.

Disturbances such as astronauts falling in must be functions of \(r,t,\tht,\) and \(\vv\)\,, and thus
generate the same disturbances in region \(II\)  as in region \(I\) , so that the clone conditions
stay exactly applicable.

\newsecl{Conclusion}{conc}
The fact that in our earlier paper we had considered using the antipodes, betrays that
we were not thinking of cloning. But now we see that involving the antipodal mapping
actually enhanced the difficulties, and we realise that nothing is more natural than assuming
the states on the points \((x,y,\W)\)
) in the Kruskal-Szekeres coordinates to be exact
quantum clones of the points \((-x,-y,\W\).  At the same time we had the need to consider
Hermitian conjugation, so that we may rely on the \(CPT\) theorem of particle physics.

By restricting the wave functions  \(\psi_{\ell,m,\,\inn}(u^+) \) to be real, our algebra, 
section \ref{harm.sec}, ensures
that  \(\psi_{\ell,m,\,\outt}(u^-)\) is real as well, so that unitarity applies.

The regions \(I\) and \(II\) of the Penrose diagram are not only identical when there is no
matter around, they will always be identical, simply because we can represent them in
identical coordinates: the original Schwarzschild frame, \((r,t,\W)\)\,. It is easy to conjecture
that this is actually a condition that may be imposed on any kind of space-time where
black holes play a role, such as black hole mergers, but we have not proven this conjecture;
the dynamical laws would be obtained by using (the generalisations of) Kruskal-Szekeres
coordinates. Only in these coordinates we see how the equations transform at points near
to \(r = 2GM\). It is just the quantum states that can be characterised more precisely in
the original Schwarzschild coordinate frame. In this frame, the two spaces will always
be clones of one another, and there are no physical effects that can disrupt the equation
that enforces the quantum clone states to continue to exactly coincide during the entire
evolution.

Had we replaced region \(II\) by what it is at its \emph{antipodes,} then the evolution law for the
two clones would not have been identical anymore (there could have been an astronaut at
one point, but empty space at his antipode), so that the states are no longer exact clones,
and the mathematics would fail.

Figure 1 shows that the regions \(III\) and \(IV\) \emph{play no role} in the evolution at all.
The Cauchy surface is indicated there with a grey line. The clones move along with the
dashed grey line. This line always pivots around the origin, so that \(III\) and \(IV\) are
avoided. This is why we say that the black hole has no interior; regions \(III\) and \(IV\) are
to be regarded as analytic extensions such as the analytic continuation towards complex
coordinates, which are very useful for solving mathematical equations, but do not have
any direct physical interpretation.

If one wants to talk about the black hole interior, one may consider region \(II\) as
the interior, but we add to this that the interior contains nothing but clones of the real
physical variables.

We reduced all physics to be described by the distribution of states on the positive
half of \(u^+\) space, which is mapped onto the positive part of \(u^-\) space through eqs. \eqn{unit3} -- \eqn{unit7}). 
This is a unitary transformation, so that it can be inverted, to see that the two
representations \eqn{unit3} and \eqn{unit5} are equivalent.

The fact that our procedure, using cloning, restores unitarity, so that no information
is lost at all, came as a surprise to us.

However, our problem was that we wish to deduce the entire evolution equation that
tells us how all Standard Model particles and gravitons are processed from in- to out-particles.
All we have done is see how it works when we know the momentum densities
or the quantum coordinates of the in-particles, to express these into the coordinates of
the out-particles. How to do the mapping to SM variables is still a technically highly
demanding problem, and that, as yet, has to be left for the future.

Finally, we conjecture that the need to limit ourselves to real wave functions rather
than complex ones, is related to the demand that, in the presence of black holes, global
additively conserved charges cannot exist, as they would lead to a denumerable infinity of
black hole states. This would violate the Bekenstein bound\,\cite{Bekenstein} for the black hole entropy.
Local conserved symmetries are certainly allowed, so that we can have conserved electric
(and weak and coloured) charges.

\end{document}